\documentclass{PoS}

\newcommand{\bea}{\begin{eqnarray}}
\newcommand{\eea}{\end{eqnarray}}

\title{A simple model with $\mathbb{Z}_{N}$ symmetry}

\ShortTitle{A simple model with $\mathbb{Z}_{N}$ symmetry}

\author{\speaker{Masanobu Yahiro}\\
        Department of Physics, Kyushu University,
             Fukuoka 812-8581, Japan\\
        E-mail: \email{yahiro@phys.kyushu-u.ac.jp}}

\author{Hiroaki Kouno\\
        Department of Physics, Saga University,
             Saga 840-8502, Japan\\
             E-mail: \email{kounoh@cc.saga-u.ac.jp}}
        
\author{Yuji Sakai\\
Quantum Hadron Physics Laboratory RIKEN Nishina Center, Saitama 351-0198, 
Japan\\E-mail: \email{ysakai@riken.jp}}

\author{Takahiro Sasaki\\
Department of Physics, Kyushu University,
             Fukuoka 812-8581, Japan\\
             E-mail: \email{sasaki@phys.kyushu-u.ac.jp}}

\author{Takahiro Makiyama\\
        Department of Physics, Saga University,
             Saga 840-8502, Japan\\
        E-mail: \email{12634019@edu.cc.saga-u.ac.jp}}
        
\abstract{
We propose a simple model with the $\mathbb{Z}_{N}$ symmetry 
in order to answer whether the symmetry is a good concept in 
QCD with light quark mass. The model is constructed by 
imposing the flavor-dependent twisted boundary condition (TBC) 
on the three-flavor Polyakov-loop extended Nambu-Jona-Lasinio model. 
In the model, the $\mathbb{Z}_{N}$ symmetry is preserved 
below some temperature $T_c$, but spontaneously broken above $T_c$. 
Dynamics of the simple model is 
similar to that of the original PNJL model without the TBC, 
indicating that the $\mathbb{Z}_{N}$ symmetry is a good concept. 
We also investigate the interplay between the $\mathbb{Z}_{N}$ symmetry and 
the emergence of the quarkyonic phase. 
}

\FullConference{The 30 International Symposium on Lattice Field Theory - Lattice 2012,\\
		June 24-29, 2012\\
		Cairns, Australia}

\begin{document}

\section{Introduction} 
In the limit of zero current quark mass, the chiral condensate 
is an exact order parameter for the chiral restoration. 
In the limit of infinite current quark mass, on the contrary,  
the Polyakov loop becomes an exact order parameter 
for the deconfinement transition, since the $\mathbb{Z}_{N}$ symmetry 
is exact there. 
For the real world in which $u$ and $d$ quarks have small current 
quark masses, 
the chiral condensate is considered to be a good order parameter, but 
it is not clear whether the Polyakov loop is a good order parameter.
In this paper, we approach this problem by proposing a simple model 
with the $\mathbb{Z}_{N}$ symmetry. 
This paper is based on our recent papers~\cite{Kouno:2012zz,Sakai:2012ika}.

We start with the SU(3) gauge theory with three degenerate flavor quarks. 
The partition function $Z$ in Euclidean spacetime is  
\bea
Z=\int Dq D\bar{q} DA \exp[-S_0] 
\label{QCD-Z}
\eea
with the action
\bea
S_0=\int d^4x [\sum_{f}\bar{q}_f(\gamma_\nu D_\nu +m_f)q_f
+{1\over{4g^2}}{F_{\mu\nu}^{a}}^2] 
\label{QCD-S}
\eea
and the temporal boundary condition 
\bea
q_f(x, \beta=1/T )=-q_f(x, 0). 
\label{period-QCD}
\eea
The $\mathbb{Z}_{3}$ transformation
changes the ferimon boundary condition as~\cite{RW,Sakai}
\bea
q_f(x, \beta)=-\exp{(i 2\pi k/{3})}q_f(x, 0)  
\label{period-QCD-Z}
\eea
for integers $k=0,1,2$, 
while the action $S_0$ keeps the original form (\ref{QCD-S}) 
since the $\mathbb{Z}_{3}$ symmetry is 
the center symmetry of the gauge symmetry~\cite{RW}. 
The $\mathbb{Z}_{3}$ symmetry thus breaks down 
through the fermion boundary condition. 
Now we assume the twisted boundary conditions (TBC) 
\begin{eqnarray}
q_f(x, \beta )&=&-\exp{(-i\theta_f)}q_f(x,0)
\equiv -\exp{[-i(\theta_1+2\pi (f-1)/3)]}q_f(x, 0)
\label{period}
\end{eqnarray}
for flavors $f=1,2,3$, where $\theta_1$ is an arbitrary real number 
in the range $0 \le \theta_1 < 2\pi$. 
QCD with the TBC has the $\mathbb{Z}_{3}$ symmetry. 
Actually the $\mathbb{Z}_{3}$ transformation changes $f$ into $f-k$, 
but $f-k$ can be relabeled by $f$ since $S_0$ is invariant under 
the relabeling. 
The TBC is useful to understand the color confinement.

When the fermion field $q_f$ is transformed as 
\begin{eqnarray}
q_f = \exp{(-i\theta_fT\tau )}q_f' 
\label{transform_1}
\end{eqnarray}
for Euclidean time $\tau$, 
the action $S_0$ is changed into 
\bea
S(\theta_f)=\int d^4x [\sum_{f}\bar{q}_f'
(\gamma_\nu D_\nu - \mu_f\gamma_4+m_f)q_f'
+{1\over{4g^2}}F_{\mu\nu}^2] 
\label{QCD1}
\eea
with the imaginary quark number chemical potential $\mu_f=i T \theta_f$, 
while  the TBC is transformed back to the standard one (\ref{period-QCD}). 
The action $S_0$ with the TBC is thus equivalent to 
the action $S(\theta_f)$ with the standard one (\ref{period-QCD}). 
The partition function $Z(T,\theta)$ 
has the Roberge-Weiss (RW) periodicity~\cite{RW}: 
$Z(T,\theta)=Z(T,\theta +2\pi k/3)$ for any integer $k$. 
The Polyakov-loop extended Nambu-Jona-Lasinio (PNJL) model 
is a good model to understand QCD at finite imaginary chemical 
potential~\cite{Sakai}.

In this paper, we consider QCD with the TBC in order to 
answer whether the $\mathbb{Z}_{N}$ symmetry is 
a good concept in QCD with light quark mass. 
Dynamics of the theory is studied concretely by 
imposing the TBC on the PNJL model, i.e., the TBC model.

\section{PNJL and TBC models}
\label{Sec:model}

The three-flavor PNJL Lagrangian is  
\bea
 {\cal L}  
&=& \sum_f {\bar q}_f(\gamma_\nu D_\nu - \mu_f \gamma_4  + m_f )q_f  
- G_{\rm S} \sum_f \sum_{a=0}^{8} 
    [({\bar q}_f \lambda_a q_f )^2 +({\bar q }_fi\gamma_5 \lambda_a q_f )^2] 
\nonumber\\
 &&+ G_{\rm D} \Bigl[\det_{ij} {\bar q}_i (1+\gamma_5) q_j 
           +\det_{ij} {\bar q}_i (1-\gamma_5) q_j \Bigr]
+{\cal U}(\Phi [A],\Phi^* [A],T) . 
\label{L_nc3}
\eea 
For the Polyakov potential ${\cal U}$ as a function 
of the Polyakov-loop $\Phi$ 
and its conjugate $\Phi ^*$, we take the  potential of Ref.~\cite{Rossner}. 
Now we consider the PNJL model with the TBC (\ref{period}), that is, 
the TBC model. 
The thermodynamic potential of the TBC model is nothing but 
that of the PNJL model with 
the flavor-dependent imaginary chemical potential $\mu_f=i\theta_fT$. 
In the mean-field level, the thermodynamic potential $\Omega$ 
of the TBC model is obtained as 
\bea
\Omega
= -&2& \sum_{f=u,d,s} \sum_{c=r,g,b}\int \frac{d^3 p}{(2\pi)^3}
   \Bigl[ E_f 
        + \frac{1}{\beta}\ln~ [1 + e^{i\phi_c}e^{i\theta_f} e^{-\beta E_f}]
        + \frac{1}{\beta}\ln~ [1 + e^{-i\phi_c}e^{-i\theta_f}e^{-\beta E_f}]
\Bigl]
\nonumber\\
&+& U(\sigma_u,\sigma_d,\sigma_s) +{\cal U}(\Phi,\Phi^*,T) 
\label{PNJL-Omega_original}, 
\eea
where $\sigma_{f} \equiv \langle {\bar q}_f q_f \rangle$, 
$E_f \equiv \sqrt{{\bf p}^3+{M_{f}}^2}$ with 
\bea
M_{f}=m_{f}-4G_{\rm S}\sigma_{f}+  2G_{\rm D} \sigma_{f^\prime } \sigma_{f^{\prime\prime}} 
\eea
for $f \neq f^\prime$ and $f \neq f^{\prime\prime}$ and $f^{\prime\prime}\neq f^{\prime\prime\prime}$. The mesonic potential 
$U(\sigma_u,\sigma_d,\sigma_s)$ are obtained by 
\bea
U(\sigma_u,\sigma_d,\sigma_s)=\sum_{f=u,d,s}  2 G_{\rm S} \sigma_{f}^2 
 - 4 G_{\rm D} \sigma_{u}\sigma_{d}\sigma_{s}. 
\label{Upotential}
\eea
The vacuum term in (\ref{PNJL-Omega_original}) is regularized with 
the three-dimensional cutoff $\Lambda$. 
For the parameter set ($G_{\rm S}$, $G_{\rm D}$, $m_l$, $m_s$, $\Lambda$), 
we take the set of Ref.~\cite{Rehberg}, except that 
the s-quark mass $m_s$ is taken to be the same as the light quark mass 
$m_l \equiv m_u=m_d$.

Taking the color summation in (\ref{PNJL-Omega_original}), we can get 
\bea
\Omega
= -2 \sum_{f=u,d,s} \int \frac{d^3 p}{(2\pi)^3}
   \Bigl[ 
   N_{\mathrm{c}} E_f 
        + &&\frac{1}{\beta}\ln [1 + C_{3,1}({\bf p}) e^{i\theta_f}
+C_{3,2}({\bf p})e^{2i\theta_f}+ C_{3,3}({\bf p})e^{3i\theta_f}]
\nonumber\\
        + &&\frac{1}{\beta}\ln [1 + 
C_{3,1}^*({\bf p})e^{-i\theta_f}e^{-\beta E_f} 
+C_{3,2}^*({\bf p})e^{-2i\theta_f}+ C_{3,3}^*({\bf p})e^{-3i\theta_f}]
	      \Bigl]
\nonumber\\
+ && U(\sigma_u,\sigma_d,\sigma_s) +{\cal U}(\Phi,\Phi^*,T) 
\label{PNJL-Omega}
\eea
with 
\begin{eqnarray}
C_{3,1}({\bf p})=3\Phi e^{-\beta E_f},~~~
C_{3,2}({\bf p})=3\Phi^* e^{-2\beta E_f},~~~
C_{3,3}({\bf p})=e^{-3\beta E_f}.
\label{factor_3_omega}
\end{eqnarray}
When $\Phi=0$, $\Omega$ has no flavor dependence, since 
$C_{3,1}=C_{3,2}=0$ and 
the factors $e^{\pm 3i \theta_f}$ do not depend on flavor. 
The flavor symmetry is thus preserved in the confinement phase with $\Phi=0$.

\section{Numerical results}

\begin{figure}[htbp]
\begin{center}
\hspace{-10pt}
\includegraphics[width=0.3\textwidth,angle=-90]{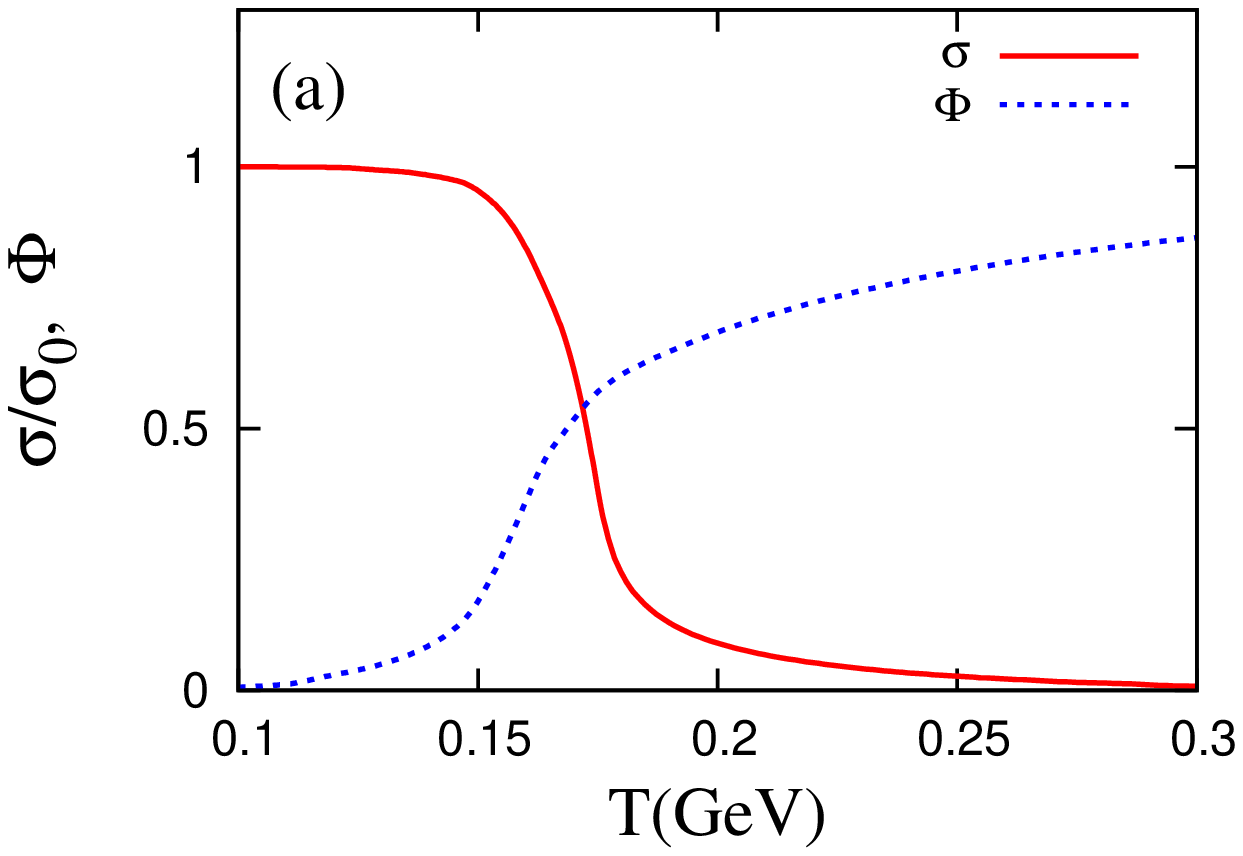}
\includegraphics[width=0.3\textwidth,angle=-90]{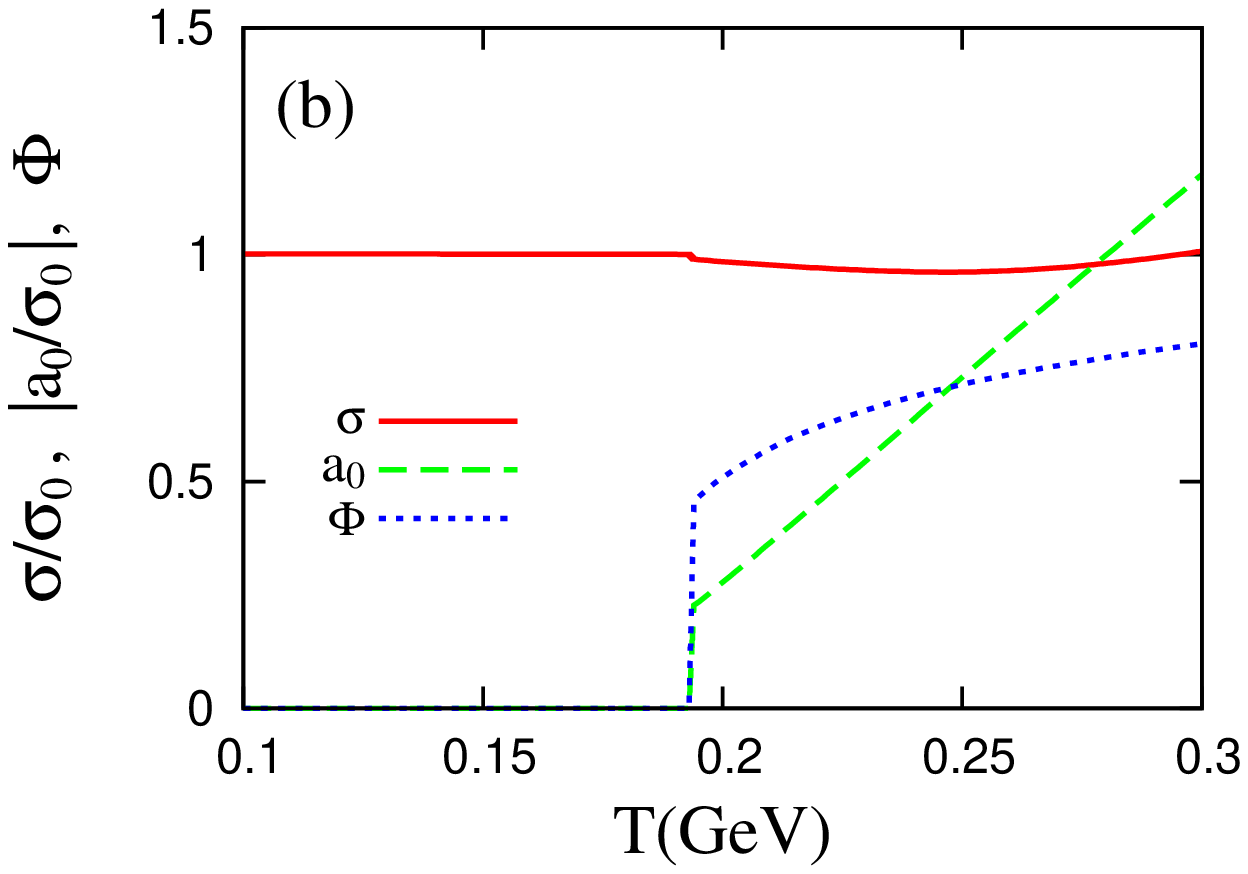}
\end{center}
\vspace{-10pt}
\caption{
$T$ dependence of $\sigma$, $\Phi$ and $a_0$ for (a) 
$(\theta_u,\theta_d,\theta_s)=(0,0,0)$ and (b) 
$(\theta_u,\theta_d,\theta_s)=(0, 2\pi /3,4\pi/3)$. 
$\sigma$ is normalized by the value $\sigma_0$ at $T=0$.  
See legends for the definition of lines.  
}
\label{nc3_mu0_order}
\end{figure}

Figure \ref{nc3_mu0_order} shows $T$ dependence of 
order parameters $\sigma$, $\Phi$ and 
$a_0\equiv  \sigma_u-\sigma_d=\sigma_u-\sigma_s$ 
in (a) the PNJL model with 
$(\theta_u,\theta_d,\theta_s)=(0,\theta,-\theta)=(0,0,0)$ and 
(b) the TBC model with 
$(\theta_u,\theta_d,\theta_s)=(0,\theta,-\theta)=(0, 2\pi /3,-2\pi/3)$; 
note that $a_0$ is an order parameter of the flavor symmetry. 
In the PNJL model, both the chiral restoration 
and the deconfinement transition are crossover. 
In the TBC model, 
the first-order deconfinement transition takes place 
at $T=T_c \approx 195$~MeV. 
Below $T_c$, $a_0$ and $\Phi$ are zero, as expected. 
The flavor symmetry is thus preserved by the color confinement. 
Above $T_c$, $a_0$ and $\Phi$ become finite, 
indicating that the flavor and $\mathbb{Z}_3$ symmetries break 
simultaneously. The chiral restoration is very slow in the TBC model 
because of the breaking of flavor symmetry.  
As mentioned above, the deconfinement transition is first-order 
at $\theta=2\pi /3$, but crossover at $\theta=0$. 
This means that there appears a crtical endpoint at some $\theta$ when 
$\theta$ is varied from $2\pi /3$ to 0.

Next we consider the entanglement-PNJL (EPNJL) 
model~\cite{Sakai5,Sasaki-T_Nf3}. 
The four-quark vertex $G_{\rm S}$ is originated in a gluon exchange 
between quarks and its higher-order diagrams. 
If the gluon field $A_{\nu}$ has a vacuum expectation value 
$\langle A_{0} \rangle$, 
$A_{\nu}$ is coupled to $\langle A_{0} \rangle$  and hence 
to $\Phi$~\cite{Kondo}. 
This effect allows $G_{\rm S}$ to depend on $\Phi$: 
$G_{\rm S}=G_{\rm S}(\Phi)$~\cite{Kondo}. 
In this paper, we simply assume the following $G_{\rm S}(\Phi )$ 
by respecting the chiral symmetry, the charge-conjugation 
symmetry~\cite{Kouno} and the extended $\mathbb{Z}_3$ symmetry~\cite{Sakai}: 
\begin{eqnarray}
G_{\rm S}(\Phi)=G_{\rm S}[1-\alpha_1\Phi\Phi^*-\alpha_2(\Phi^3+\Phi^{*3})]. 
\label{entanglement-vertex}
\end{eqnarray}
In principle, $G_{\rm D}$ can also depend on $\Phi$. 
However, $\Phi$-dependence of $G_{\rm D}$ 
yields qualitatively the same effect on the phase diagram as that of 
$G_{\rm S}$~\cite{Sasaki-T_Nf3}. We can then neglect 
$\Phi$-dependence of $G_{\rm D}$. 
The parameters $\alpha_1$ and $\alpha_2$ in (\ref{entanglement-vertex}) 
are so determined as to reproduce two results of LQCD at finite $T$; 
one is the result of 2+1 flavor LQCD at $\mu=0$~\cite{YAoki} 
that the chiral transition is crossover at the physical point and 
another is the result of degenerate three-flavor LQCD 
at $\theta=\pi$~\cite{FP2010} that the order of the RW endpoint is 
first-order for small and large quark masses but second-order 
for intermediate quark masses. 
The parameter set $(\alpha_1, \alpha_2)$ satisfying these conditions 
is located in the triangle region~\cite{Sasaki-T_Nf3}
\bea
\{-1.5\alpha_1+0.3 < \alpha_2 <-0.86\alpha_1+0.32,~\alpha_2 >0\}. 
\label{triangle}
\eea
As a typical example, we take $\alpha_1=0.25$ and $\alpha_2=0.1$,  
following Ref.~\cite{Sasaki-T_Nf3}.

\begin{figure}[htbp]
\begin{center}
\hspace{-10pt}
\includegraphics[width=0.3\textwidth,angle=-90]{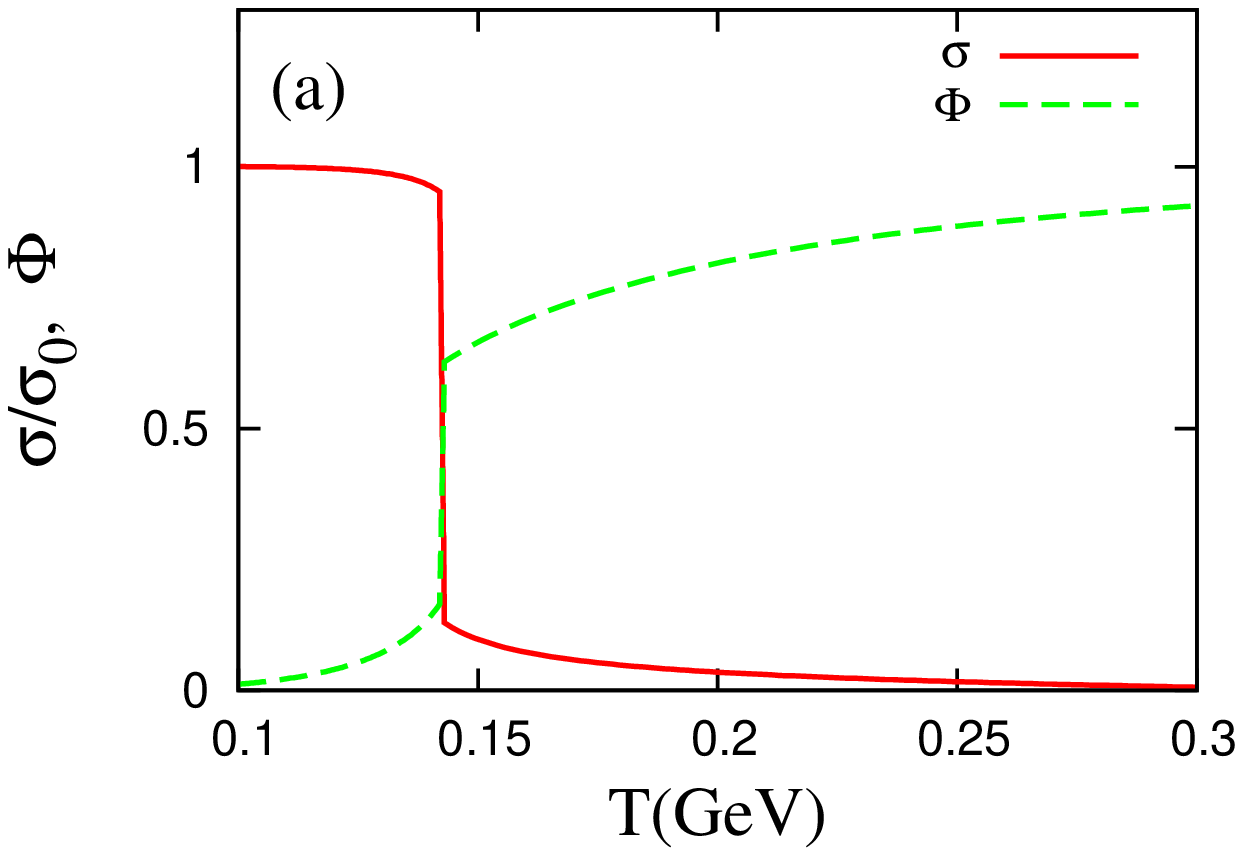}
\includegraphics[width=0.3\textwidth,angle=-90]{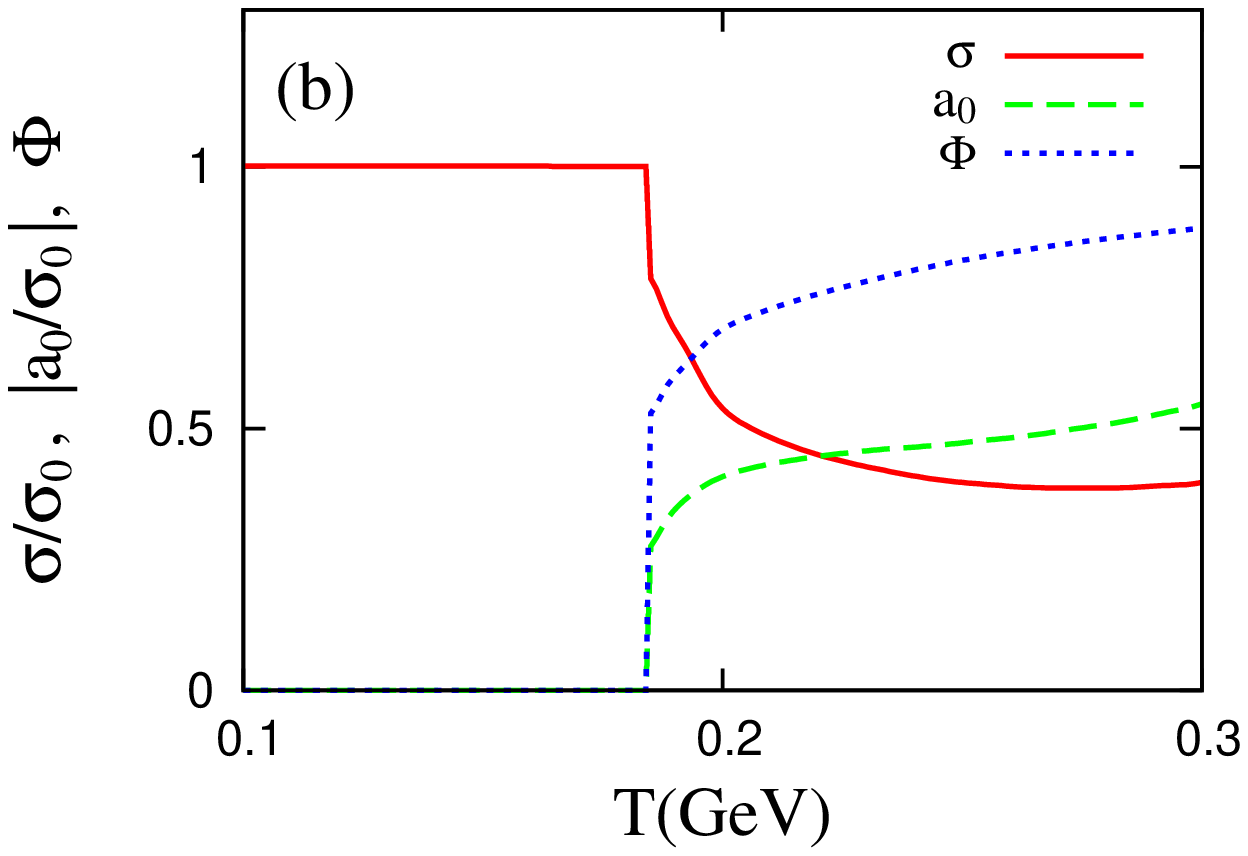}
\end{center}
\vspace{-10pt}
\caption{$T$ dependence of order parameters 
$\sigma$, $a_0$ and $\Phi$ 
calculated with the EPNJL model 
for (a) $(\theta_u,\theta_d,\theta_s)=(0,0,0)$ and 
(b) $(\theta_u,\theta_d,\theta_s)=(0,2\pi /3,4\pi/3$). 
Here $\sigma$ and $a_0$ are normalized by $\sigma_0$. 
Note that $a_0=0$ in panel (a) and $a_0\ge 0$ in panel (b), 
while $\sigma <0$ in both panels. See legends for the definition of lines.  
}
\label{nc3_mu0_order_epnjl}
\end{figure}

In Fig. \ref{nc3_mu0_order_epnjl}, $\sigma$, $a_0$ and $\Phi$ 
are calculated as a function of $T$ with the EPNJL model 
for (a) $(\theta_u,\theta_d,\theta_s)=(0,0,0)$ and (b) 
$(\theta_u,\theta_d,\theta_s)=(0,2\pi /3,4\pi/3$). 
In panel (a), 
the chiral restoration and the deconfinement transition 
are first-order, because the current quark mass ($5.5$MeV) is small 
and the correlation between $\sigma_f$ and $\Phi$ 
is strong~\cite{Sasaki-T_Nf3}. 
In panel (b), one can see that $\Phi=a_0=0$ in the confinement phase. 
The EPNJL model with the TBC yields similar $T$ 
dependence to that without the TBC 
for both the chiral restoration and the deconfinement transition, since 
the flavor-symmetry breaking above $T_c$ is weakened by 
the strong correlation between $\sigma_f$ and $\Phi$.

Now we consider the PNJL model with the flavor-dependent complex 
chemical potentials $\mu_f = \mu+i T \theta_f$ in which 
the $\theta_f$ are defined by 
\bea
(\theta_f)=(0,\theta,-\theta) . 
\label{NBC}
\eea
The present model with the $\mu_f$ is reduced to the standard PNJL model 
with the flavor-independent  real chemical potential $\mu$ 
when $\theta=0$ and 
to the TBC model with real $\mu$ when $\theta=2\pi/3$. 
Varying $\theta$ from 0 to $\theta=2\pi/3$, 
one can see how the phase diagram changes 
between the approximate color-confinement 
in the standard PNJL model and the exact color-confinement in the TBC model.

\begin{figure}[htbp]
\begin{center}
\includegraphics[width=0.3\textwidth,angle=-90]{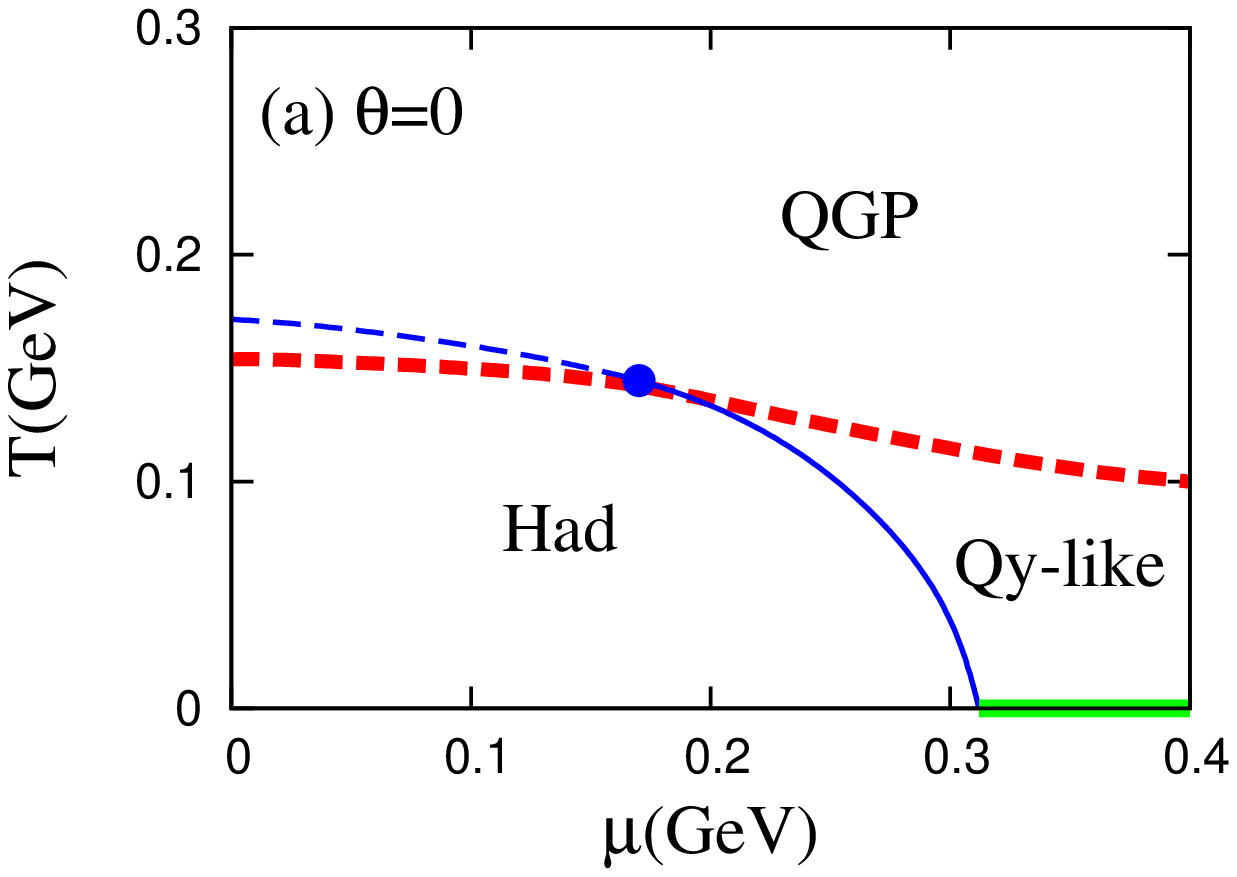}
\includegraphics[width=0.3\textwidth,angle=-90]{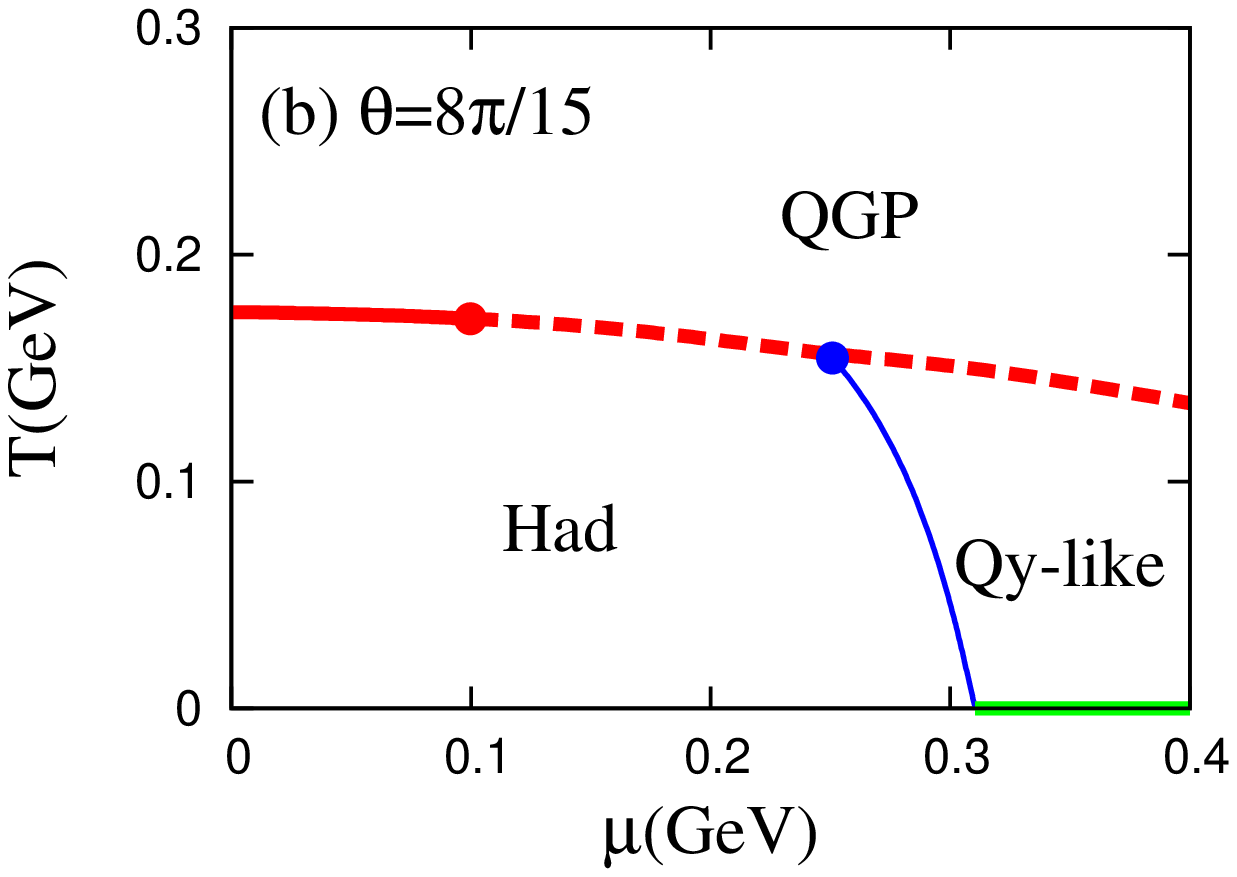}
\includegraphics[width=0.3\textwidth,angle=-90]{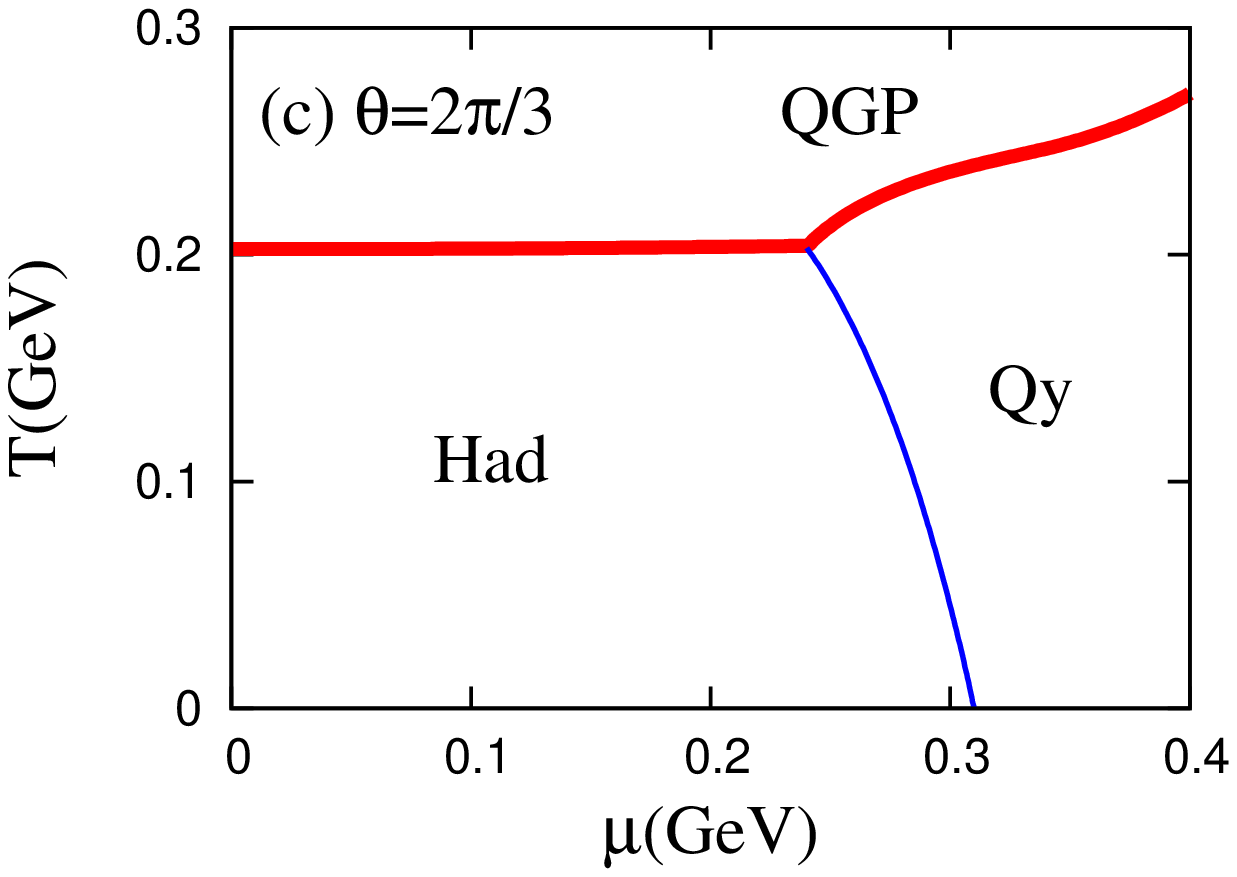}
\end{center}
\caption{
Phase diagram in the $\mu$-$T$ plane.
Panels (a)-(c) correspond to three cases 
of $\theta=0$, $8\pi/15$ and $2\pi/3$, respectively. 
The thick (thin) solid curve means the first-order deconfinement (chiral) 
phase transition line, 
while the thick (thin) dashed curve does the deconfinement (chiral) 
crossover line. The closed circles stand for the endpoints of the 
first-order deconfinement and chiral phase transition lines.
In panels (a) and (b), 
the thick-solid line at $T=0$ and $\mu \gtrsim M_f=323$~MeV 
represents the quarkyonic phase. 
}
\label{PD}
\end{figure}

Figure \ref{PD} shows the phase diagram in the $\mu$-$T$ plane. 
Panels (a)-(c) correspond to three cases 
of $\theta=0$, $8\pi/15$ and $2\pi/3$, respectively. 
When $\theta=0$, 
both the chiral and deconfinement transitions are crossover
at smaller $\mu$, but the chiral transition becomes first-order 
at larger $\mu$.
When $\theta=2\pi/3$, the deconfinement transition is the 
first-order at any $\mu$, 
whereas the chiral transition line becomes first-order only 
at $\mu \approx M_f=323$~MeV.
The region labeled by ``Qy" at $\mu \gtrsim M_f$ and small $T$ is 
the quarkyonic phase~\cite{McLerran1},  
since $\Phi=0$ but the quark number density $n$ is finite there. 
The region labeled by ``Had" is the hadron phase, because 
the chiral symmetry is broken there and thereby the equation of state is 
dominated by the pion gas~\cite{Sakai_hadron}. 
The region labeled by ``QGP" corresponds to the quark gluon plasma (QGP) 
phase, although the flavor symmetry is broken there by the TBC. 
As $\theta$ decreases from $2\pi/3$ to zero, 
the first-order chiral transition line declines toward smaller $\mu$ and 
the critical endpoint moves to smaller $\mu$.
Once $\theta$ varies from $2\pi/3$, 
the quarkyonic phase defined by $\Phi=0$ and $n \neq 0$
shrinks on a line with $T=0$ and $\mu \gtrsim M_f$ and 
a region at small $T$ and $\mu \gtrsim M_f$ becomes 
a quarkyonic-like phase with small but finite $\Phi$ and $n \ne 0$; 
the latter region is labeled by ``Qy-like".

\section{Summary}
\label{Summary}
We have proposed a simple model with the $\mathbb{Z}_{N}$ symmetry 
in order to answer whether the $\mathbb{Z}_{N}$ symmetry is a good concept 
in QCD with light quark mass. 
The model called the TBC model is constructed by imposing 
the flavor-dependent twisted boundary condition (\ref{period}) on 
the PNJL model.

In the TBC model, the $\mathbb{Z}_{3}$ symmetry is preserved 
below $T_c$, but spontaneously broken above $T_c$. 
Below $T_c$, the color confinement preserves the flavor symmetry. 
Above $T_c$, meanwhile, the flavor symmetry is broken explicitly by 
the TBC. 
The flavor-symmetry breaking makes the chiral restoration slower, but 
the entanglement interaction between $\sigma$ and $\Phi$ makes 
the restoration faster. 
Consequently, we can except that QCD with the TBC is similar to 
original QCD with the standard quark boundary condition.

We have also investigated 
the interplay between the $\mathbb{Z}_{N}$ symmetry and the emergence of 
the quarkyonic phase, considering the complex chemical potentials 
$\mu_f=\mu+i T \theta_f$ with $(\theta_f)=(0,\theta,-\theta)$ in 
the PNJL model. The PNJL model with the $\mu_f$ is reduced 
to the PNJL model with real $\mu$ for $\theta=0$, but 
to the TBC model with real $\mu$ for $\theta=2\pi/3$. 
When $\theta=2\pi/3$, the quarkyonic phase defined by $\Phi=0$ and $n > 0$ 
really exists at small $T$ and large $\mu$. 
Once $\theta$ varies from $2\pi/3$ to zero, 
the $\mathbb{Z}_{N_c}$ symmetry is broken and 
thereby the quarkyonic phase exists only 
on a line of $T=0$ and $\mu \gtrsim M_f$. The region 
at small $T$ and large $\mu$ is dominated by 
the quarkyonic-like phase characterized by small but finite $\Phi$ and $n>0$. 
The quarkyonic-like phase at $\theta=0$ is thus 
a remnant of the quarkyonic phase at $\theta=2\pi/3$.  
Since the $\mathbb{Z}_{N}$ symmetry is explicitly broken at 
$\theta=0$, it is natural to expand the concept of 
the quarkyonic phase and redefine it by a phase with small $\Phi$ and 
finite $n$. For this reason, the quarkyonic-like phase is 
often called the quarkyonic phase. 
The gross structure of the phase diagram thus has no qualitative 
difference between $\theta=2\pi/3$ and zero, 
if the concept of the quarkyonic phase is properly expanded. 
In this sense, the $\mathbb{Z}_{3}$ symmetry is a good approximate concept 
for the case of $\theta=0$, even if the current quark mass is small.


\begin{thebibliography}{99}

\expandafter\ifx\csname natexlab\endcsname\relax\def\natexlab#1{#1}\fi
\expandafter\ifx\csname bibnamefont\endcsname\relax
  \def\bibnamefont#1{#1}\fi
\expandafter\ifx\csname bibfnamefont\endcsname\relax
  \def\bibfnamefont#1{#1}\fi
\expandafter\ifx\csname citenamefont\endcsname\relax
  \def\citenamefont#1{#1}\fi
\expandafter\ifx\csname url\endcsname\relax
  \def\url#1{\texttt{#1}}\fi
\expandafter\ifx\csname urlprefix\endcsname\relax\def\urlprefix{URL }\fi
\providecommand{\bibinfo}[2]{#2}
\providecommand{\eprint}[2][]{\url{#2}}
%
%
\bibitem{Kouno:2012zz} 
  H.~Kouno, Y.~Sakai, T.~Makiyama, K.~Tokunaga, T.~Sasaki and M.~Yahiro,
  J.\ Phys.\ G {\bf 39}, 085010 (2012).
\bibitem{Sakai:2012ika} 
  Y.~Sakai, H.~Kouno, T.~Sasaki and M.~Yahiro,
  to be published in Phys. Lett. B [arXiv:1204.0228 [hep-ph]].  
\bibitem{RW}
\bibinfo{author}{\bibfnamefont{A.}~\bibnamefont{Roberge}} \bibnamefont{and}
\bibinfo{author}{\bibfnamefont{N.}~\bibnamefont{Weiss}},  
\bibinfo{journal}{Nucl. Phys. } \textbf{\bibinfo{volume}{B275}},
\bibinfo{pages}{734} (\bibinfo{year}{1986}). 
%
\bibitem{Sakai}
\bibinfo{author}{\bibfnamefont{Y.}~\bibnamefont{Sakai}},
\bibinfo{author}{\bibfnamefont{K.}~\bibnamefont{Kashiwa}}, 
\bibinfo{author}{\bibfnamefont{H.}~\bibnamefont{Kouno}}, 
\bibnamefont{and}
\bibinfo{author}{\bibfnamefont{M.}~\bibnamefont{Yahiro}},
\bibinfo{journal}{Phys.\ Rev.\  D} \textbf{\bibinfo{volume}{77}},
\bibinfo{pages}{051901(R)} (\bibinfo{year}{2008});
\bibinfo{journal}{Phys.\ Rev.\  D} \textbf{\bibinfo{volume}{78}},
\bibinfo{pages}{ 036001} (\bibinfo{year}{2008}) and references therein. 
%
%
\bibitem{Rossner}
\bibinfo{author}{\bibfnamefont{S.}~\bibnamefont{R\"{o}{\ss}ner}},
\bibinfo{author}{\bibfnamefont{C.}~\bibnamefont{Ratti}},
\bibnamefont{and}
\bibinfo{author}{\bibfnamefont{W.}~\bibnamefont{Weise}},  
  \bibinfo{journal}{Phys. Rev.\ D} \textbf{\bibinfo{volume}{75}},
  \bibinfo{pages}{034007} (\bibinfo{year}{2007}). 
%
\bibitem{Rehberg}
\bibinfo{author}{\bibfnamefont{P.}~\bibnamefont{Rehberg}},
\bibinfo{author}{\bibfnamefont{S.P.}~\bibnamefont{Klevansky}}
\bibnamefont{and}
\bibinfo{author}{\bibfnamefont{J.}~\bibnamefont{H\"{u}fner}}, 
  \bibinfo{journal}{Phys.\ Rev.\  C} \textbf{\bibinfo{volume}{53}},
  \bibinfo{pages}{410} (\bibinfo{year}{1996}). 
%
\bibitem{Sakai5}
\bibinfo{author}{\bibfnamefont{Y.}~\bibnamefont{Sakai}},
\bibinfo{author}{\bibfnamefont{T.}~\bibnamefont{Sasaki}}, 
\bibinfo{author}{\bibfnamefont{H.}~\bibnamefont{Kouno}},
\bibnamefont{and}
\bibinfo{author}{\bibfnamefont{M.}~\bibnamefont{Yahiro}}, 
\bibinfo{journal}{Phys. Rev. \ D} 
\textbf{\bibinfo{volume}{82}},
\bibinfo{pages}{076003} (\bibinfo{year}{2010}). 
%
\bibitem{Sasaki-T_Nf3}
\bibinfo{author}{\bibfnamefont{T.}~\bibnamefont{Sasaki}}, 
\bibinfo{author}{\bibfnamefont{Y.}~\bibnamefont{Sakai}}, 
\bibinfo{author}{\bibfnamefont{H.}~\bibnamefont{Kouno}}, 
\bibnamefont{and}
\bibinfo{author}{\bibfnamefont{M.}~\bibnamefont{Yahiro}}, 
\bibinfo{journal}{Phys. Rev. \ D} 
\textbf{\bibinfo{volume}{84}},
\bibinfo{pages}{091901} (\bibinfo{year}{2011}); 
%
\bibitem{Kondo}
\bibinfo{author}{\bibfnamefont{K.-I.}~\bibnamefont{Kondo}}, 
\bibinfo{journal}{Phys. Rev.\ D} \textbf{\bibinfo{volume}{82}},
\bibinfo{pages}{065024} (\bibinfo{year}{2010}).
%
\bibitem{Kouno}
\bibinfo{author}{\bibfnamefont{H.}~\bibnamefont{Kouno}},
\bibinfo{author}{\bibfnamefont{Y.}~\bibnamefont{Sakai}}, 
\bibinfo{author}{\bibfnamefont{K.}~\bibnamefont{Kashiwa}}, 
\bibnamefont{and}
\bibinfo{author}{\bibfnamefont{M.}~\bibnamefont{Yahiro}},
\bibinfo{journal}{J. Phys. \  G: Nucl. Part. Phys.} \textbf{\bibinfo{volume}{36}},
\bibinfo{pages}{115010} (\bibinfo{year}{2009});
\bibinfo{author}{\bibfnamefont{H.}~\bibnamefont{Kouno}}, 
\bibinfo{author}{\bibfnamefont{Y.}~\bibnamefont{Sakai}},
\bibinfo{author}{\bibfnamefont{T.}~\bibnamefont{Sasaki}}, 
\bibinfo{author}{\bibfnamefont{K.}~\bibnamefont{Kashiwa}}, 
\bibnamefont{and}
\bibinfo{author}{\bibfnamefont{M.}~\bibnamefont{Yahiro}},  
\bibinfo{journal}{Phys.\ Rev.\  D} \textbf{\bibinfo{volume}{83}},
\bibinfo{pages}{ 076009} (\bibinfo{year}{2011});
\bibinfo{author}{\bibfnamefont{H.}~\bibnamefont{Kouno}},
\bibinfo{author}{\bibfnamefont{M.}~\bibnamefont{Kishikawa}},
\bibinfo{author}{\bibfnamefont{T.}~\bibnamefont{Sasaki}}, 
\bibinfo{author}{\bibfnamefont{Y.}~\bibnamefont{Sakai}},
  \bibnamefont{and}
\bibinfo{author}{\bibfnamefont{M.}~\bibnamefont{Yahiro}}, 
\bibinfo{journal}{Phys. Rev. \ D} 
\textbf{\bibinfo{volume}{85}},
\bibinfo{pages}{016001} (\bibinfo{year}{2012}).  
%
\bibitem{YAoki}
Y. Aoki, G. Endr\"{o}di, Z. Fodor, S. D. Katz and K. K. Szab\'{o}, 
Nature {\bf 443}, 675 (2006). 
%
\bibitem{FP2010}
\bibinfo{author}{\bibfnamefont{P.}~\bibnamefont{de}~\bibnamefont
{Forcrand}} 
\bibnamefont{and}
\bibinfo{author}{\bibfnamefont{O.}~\bibnamefont{Philipsen}},  
\bibinfo{journal}{Phys. Rev. Lett.} \textbf{\bibinfo{volume}{105}},
\bibinfo{pages}{152001} (\bibinfo{year}{2010}). 
%
\bibitem{McLerran1}
\bibinfo{author}{\bibfnamefont{L.}~\bibnamefont{McLerran}},
\bibnamefont{and}
\bibinfo{author}{\bibfnamefont{R.}~\bibfnamefont{D.}~\bibnamefont{Pisarski}},
\bibinfo{journal}{Nucl. Phys.} \textbf{\bibinfo{volume}{A796}},
\bibinfo{pages}{83} (\bibinfo{year}{2007}); \\
\bibinfo{author}{\bibfnamefont{Y.}~\bibnamefont{Hidaka}},
\bibinfo{author}{\bibfnamefont{L.}~\bibnamefont{McLerran}},
\bibnamefont{and}
\bibinfo{author}{\bibfnamefont{R.}~\bibfnamefont{D.}~\bibnamefont{Pisarski}},
\bibinfo{journal}{Nucl. Phys.} \textbf{\bibinfo{volume}{A808}},
\bibinfo{pages}{117} (\bibinfo{year}{2008}). 

%
\bibitem{Sakai_hadron}
\bibinfo{author}{\bibfnamefont{Y.}~\bibnamefont{Sakai}},
\bibinfo{author}{\bibfnamefont{T.}~\bibnamefont{Sasaki}}, 
\bibinfo{author}{\bibfnamefont{H.}~\bibnamefont{Kouno}},
\bibnamefont{and}
\bibinfo{author}{\bibfnamefont{M.}~\bibnamefont{Yahiro}}, 
\bibinfo{journal}{J. Phys. \ G} 
\textbf{\bibinfo{volume}{39}},
\bibinfo{pages}{035004} (\bibinfo{year}{2012}). 


\end{thebibliography}
\end{document}